\documentclass[11pt]{article}
\usepackage{a4wide}
\usepackage{times,mathptmx}
\usepackage{amsmath}
\usepackage{verbatim,alltt}
\usepackage[hyphens]{url}
\usepackage{graphicx,alltt}

\begin{document}
\def\thepage{\arabic{page}}
\title{PAKE-based mutual HTTP authentication \\ for preventing phishing attacks}

\author{Yutaka Oiwa,\quad Hajime Watanabe,\quad Hiromitsu Takagi \\
RCIS, AIST\thanks{Research Center for Information Security (RCIS),
National Institute of Advanced Industrial Science and Technology (AIST).}
}

\maketitle

\sloppy

\abstract{This paper describes a new password-based mutual
authentication protocol for Web systems which prevents various kinds of
phishing attacks.  This protocol provides a protection of user's
passwords against any phishers even if dictionary attack is employed,
and prevents phishers from imitating a false sense of successful
authentication to users.  The protocol is designed considering
interoperability with many recent Web applications
which requires many features which current HTTP authentication does not provide.
The protocol is proposed as an Internet Draft submitted to
IETF, and implemented in both server side (as an Apache extension) and
client side (as a Mozilla-based browser and an IE-based one).
The paper also proposes a new user-interface for this protocol
which is always distinguishable from fake dialogs provided by phishers.
}

\section{Introduction}

\subsection{Summary of this paper}

This paper describes a new password-based mutual authentication protocol for Web
systems which prevents various kinds of phishing attacks.  Currently, 
initial design of the protocol is finished, an extension for Apache
Web server and an Mozilla-based extended browser supporting the new protocol
are implemented, and the specification is submitted to IETF as an Internet
Draft~\cite{draft-oiwa-http-mutualauth-05} for standardization.  This paper
describes the criteria and decisions behind the protocol design, and
the difference from previous work, which are not discussed in the specification document.

Recently, phishing attacks are getting more and more sophisticated.
Phishers not only steal user's password directly, but also imitate
successful authentication to steal user's private information, check
the password validity by forwarding the password to the legitimate
server, or employ a man-in-the-middle style attack to hijack user's login
session.  Existing countermeasures such as one-time passwords can not
completely solve these problems.

The new protocol prevents such attacks by providing users a way to
discriminate between true and fake web servers using their own
passwords.  Even when a user inputs his/her password to a fake
website, using this authentication method, any information about the
password does not leak to the phisher, and the user certainly notices
that the mutual authentication has failed.  Phishers cannot make such
authentication attempt succeed, even if they forward received data
from a user to the legitimate server or vice versa.  Users can safely
enter private data to the web forms after confirming that the mutual
authentication has succeeded.

To achieve this goal, we use a mechanism in ISO/IEC 11770-4, a kind of
PAKE (Password-Authenticated Key Exchange) authentication algorithms
as a basis.  The use of a PAKE mechanism allows users to use familiar
ID/password based accesses, without fear of leaking any password
information to the communication peer.  The protocol, as a whole, is
designed as a natural extension to the current HTTP authentication
schema such as Basic and Digest access authentication~\cite{RFC2617}. To
use PAKE mechanism for such a purpose, we had to modify it to prevent
credential forwarding attacks (man-in-the-middle attacks).  The protocol
copes with HTTPS to provide encryption, and it also
supports the use of load
balancers and SSL accelerators for easy deployment on real systems.

There are some existing work (e.g. TLS-SRP extensions) which uses a
kind of PAKE algorithm in the transport layer. Those protocols can be
used for closed applications like VPN or IPP.  However, we have
considered that it is not suitable for general web systems which
perform authentication in the application layer.

This paper also proposes new user-interface for this authentication system.
To prevent phishing attacks, it is important to make users easily
check whether the server authentication has been succeeded or not.
This information must not be forged by phishers, otherwise
phishers deliberately circumvent users that the mutual authentication is
established so that users input private information to the phishing
sites.  We have implemented an extension for Mozilla Firefox which uses the address-bar area
for password input instead of using dialogs, 
and introduces an indicator for displaying the
result of authentication.  When a user wants to input private
information, he/she just needs to check that this indicator has good
(green) indication with the user's ID.  This indication ensures that
the document displayed in the content area is not from any phishing sites.

\subsection{Phishing attacks}

Throughout this paper, we define phishing as to make a false web service which
runs on its own host and imitates itself as an existing legitimate service
running on another host, in order to steal (or alter) some information
which users intend to send to the legitimate service.
This kind of attack is easy to be done on the Internet: everyone including
a rogue person can start running their own web server, and make a visual 
imitation of any legitimate servers by simply copying their visual elements
such as logotypes.  It is a inherent weakness of the Internet which comes
from the openness of it: we can not prevent such servers from running.
This is also true for HTTPS services in general
(see also the discussion in Section~\ref{sec:globalwhitelist}).

The fundamental way of defense against phishing is to check the hostname
displayed on the address bar (or other identities such as the subjects of server certificates under HTTPS protocol)
service just before sending critical information to the server.
However, phishers often use social attack techniques so that users 
mistakenly send information to the phishing host without noticing that
the hostname is different from the genuine host.
The purpose of the protocol is to provide a secure alternative way
to check host's identity.

In the above definition, we ignored some transport-level attacks, 
such as DNS spoofing or transport-level man-in-the-middle attacks, because 
the TLS (HTTPS) protocol works well against such attacks.

\subsection{The structure of this paper}

The rest of this paper is organized as follows: in
Section~\ref{sec:background}, we will investigate current phishing
attacks and existing countermeasures for it.  Section~\ref{sec:proposal}
describes the design goals and features of the protocol.
Section~\ref{sec:protocol} contains a brief view of the protocol,
and Section~\ref{sec:ui} shows our user interface design which is an
important part of our proposal.  Section~\ref{sec:morefeatures}
describes our extension to HTTP authentication protocol to make it
friendly with existing web applications.
Our implementation and public experiments are briefly introduced in
Section~\ref{sec:implementation}.
Section~\ref{sec:discussion} has discussions on the proposal,
and examine previous work in Section~\ref{sec:relatedwork}.
Section~\ref{sec:conclusion} concludes this paper and shows future
direction of our research.

\section{Background}\label{sec:background}

\subsection{Categorization of phishing attacks}\label{sec:phishing}

\begin{figure}[t]
\center{\noindent\includegraphics[width=0.6\textwidth]{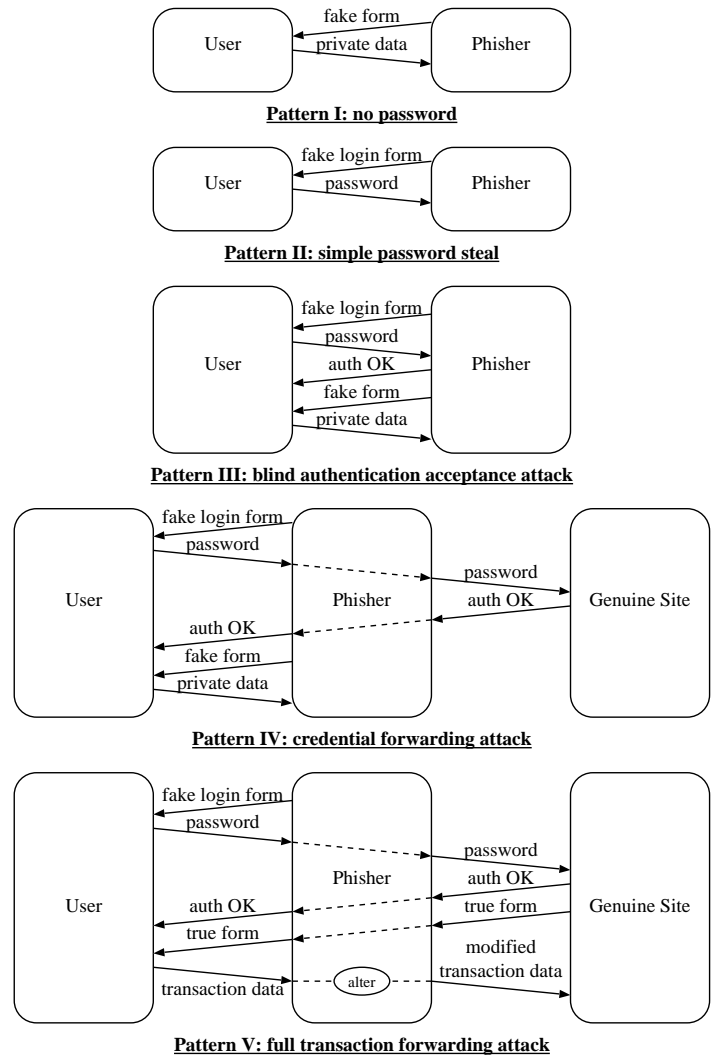}}
\caption{Phishing attack patterns}\label{fig:phishpatterns}
\end{figure}

Phishing attacks are very common in recent days, and its method is
getting tricky.  Many users including novice and even experienced users
are still trapped with kind of attacks.  Examining existing
phishing sites, we can find several patterns for how the phishing sites
handled users' passwords which are mistakenly inputted by users.  Common
patterns can be categorized as follows:
\begin{quote}
 \begin{description}
  \item[Pattern I (simple forgery without password):] Do not use and acquire passwords at all.  Instead,
	     acquire other private information such as credit card
	     numbers.
  \item[Pattern II (simple password stealing):] Ask users' password and does nothing else.
  \item[Pattern III (blind authentication acceptance):] Ask users' passwords, always accept it and
	     continue to forge users.
  \item[Pattern IV (credential forwarding):] Ask users' passwords, then use genuine sites to check their validity.
  \item[Pattern V (full transaction forwarding):] Forward all communications (including passwords) to genuine sites,
	     possibly altering part of these (e.g.,~altering bank
	     account number for money transfers to steal
	     money).
\end{description}
\end{quote}
Figure~\ref{fig:phishpatterns} summarizes those patterns.

A \textbf{pattern I attack} is simple: first instance of phishing in the early
days was to simply show imitated web forms to let victim users input
private information such as credit card numbers and PINs, without any
authentication attempts.  These simple attack scheme are still in use, however, as
users are getting accustomed to phishing attacks, more sophisticated attacks
appeared to forge users.

A \textbf{Pattern II attack}, another instance in the early days was to steal user's password.
After acquiring passwords from form inputs, phishers simply show
either communication errors, authentication errors, or just a ``thanks''
message.  We categorize this as a ``password-stealing attack.''

Later, there was new \textbf{pattern III attack}, which was an extension of pattern I 
with fake password authentication added to it.
Possible background of this phishing pattern might be that 
phishing was getting popular and people start to 
feel unnatural for the pattern II phishing sites.
When users entered the password and get no meaningful response,
they begin feeling something unusual and change their password immediately.
New kinds of phishing sites first show a fake authentication requests, and
when a user enters a password, they show another form to input
private information like pattern I attacks regardless of the inputted
passwords.  Because victim users have been inputted passwords,
they believe in that they are communicating with
genuine sites.  We call these ``blind authentication acceptance attacks''.

When the blind acceptance attacks became common, users began to take a
countermeasure to them.  To detect such attacks, users first input wrong
passwords to every site they are visiting and see the responses from the
server.  As phishing sites do not know the true passwords, their
response to wrong passwords will be ``login succeeded''.
However, this countermeasure was not perfect: some phishing sites
started sending the inputted passwords to the genuine sites to see
whether they are correct ones.  These attacks are hereafter named
``credential forwarding attacks'', or ``\textbf{pattern IV attacks}''.

Another sophisticated phishing sites ever seen are to forward almost all
web traffic to the genuine sites and to steal or hijack the web session
of the user~\cite{washingtonpost-200607-phishing}. When a user sent a
critical information such as a bank account number for money transfer,
they alter it to make false request (e.g. redirect bank money transfer
to their own accounts). We name it a ``full-transaction forwarding (\textbf{pattern V})
attack''.  This attack works even when there are additional (two-factor)
authentication present.

Note that the last pattern is completely different from transport-level
man-in-the-middle attacks.  In transport-level man-in-the-middle
attacks, the client \emph{software} is forged to communicate with
false servers.  In phishing cases, however, the client \emph{user} is
forged, and the client software is directed by user to communicate
with phishing hosts.  This makes some cryptographic solution for
man-in-the-middle attacks (such as TLS) ineffective against phishing attacks.

\subsection{Existing countermeasures against phishing}\label{sec:countermeasures}

There are many proposed countermeasures against phishing attacks.
However, none of them are satisfactory and complete.
Table~\ref{tbl:compare} summarizes the comparison in this section.
It shows, for each countermeasures, whether the measure is
effective for various patterns of phishing attacks
shown in the previous section.

\begin{table*}[t]
\caption{Comparison of Phising countermeasures}\label{tbl:compare}
\begin{center}\scriptsize
\begin{tabular}{l|ccccc|c|c|c}
& \multicolumn{5}{c|}{Prevents Phishing Patterns} & Applicable &
 Detecting & Portable \\
& I & II & III & IV & V & for all & all & (no local\\
& no pwd. & pwd. steal & accept & pwd. fwd. & session fwd. & genuine
 sites$^{\dagger1}$ & phishing sites$^{\dagger2}$ & storage)$^{\dagger3}$ \\\hline
Global whitelist & \textit{yes} & \textit{yes} & \textit{yes} & \textit{yes} & \textit{yes} & NO & -- & \textit{yes} \\
Global blacklist & \textit{yes} & \textit{yes} & \textit{yes} & \textit{yes} & \textit{yes} & -- & NO & \textit{yes} \\
Local whitelist  & \textit{yes} & \textit{yes} & \textit{yes} & \textit{yes} & \textit{yes} & NO & -- & NO \\
 SSL with client key & NO & \textit{yes} & NO & \textit{yes} & \textit{yes} & \textit{yes}? & NO & NO \\
 Digest authentication & ? & Weak$^{\dagger3}$ & NO & NO & NO & \textit{yes} & NO & \textit{yes}  \\
 PwdHash & ? & Weak$^{\dagger3}$ & NO & \textit{yes} & \textit{yes} & \textit{yes} & NO & \textit{yes} \\ \hline
 Our proposal & \textit{yes} & \textit{yes} & \textit{yes} & \textit{yes} & \textit{yes} & \textit{yes} & \textit{yes} & \textit{yes}
\end{tabular}\\
 \vspace{\baselineskip}
\end{center}
\begin{quote}\scriptsize
\begin{enumerate}
 \item[$^{\dagger1}$] Every sites, including both large-scale and small-scale
 (e.g. personal) sites, have opportunity to use the protocol without any
 social requirements (e.g. third-party screening or auditing).
 \item[$^{\dagger2}$] this means that whether phishing sites fails to imitate
		      the behavior associated to genuine sites under
		      each countermeasure (e.g. successful authentication).
		      social requirements (e.g. third-party screening or auditing).
 \item[$^{\dagger3}$] Whether users are free from managing information
		      (such as private keys) stored in client software.
 \item[$^{\dagger4}$] ``Weak'' means cryptographically weak against off-line dictionary attacks.
\end{enumerate}
\end{quote}
\end{table*}

\subsubsection{Global whitelists and blacklists}\label{sec:globalwhitelist}

Most common approaches against phishing attacks are to maintain a list
of either good or bad sites by trusted third-party.  The common
problem of those approaches are that such methods cannot be complete by
the Internet's nature.  Especially for whitelist approaches, there is
always a tradeoff between rejection of legitimate servers (false
positive) and false acceptance of phishing servers (false negative), and
it seems to be impossible to give an appropriate rights for every
``good'' or ``genuine'' website on the Internet while keeping all
phishing sites away.

In former days, SSL server certificates are thought to be providing some
kind of trusts over commerce web sites, as a kind of whitelist.  Valid
SSL certificates are only issued by designated certificate authorities
(CAs) which are trusted by Web browser vendors.  However, the validation
for issuing SSL certificates is too easy for phishers in fact.
Moreover, many CAs recently started to issue so-called ``class-1 SSL
server certificates'' or ``domain-validated certificates'' which can be
obtained without any evidence of real-world identities.  In the reality,
there are phishing sites which runs HTTPS sites with valid SSL server
certificates.  Although SSL server certificates still provide an
important role for transport security\footnote{With server certificates
issued by trusted CAs, TLS ensures that the client software is communicating to
\emph{some} server who are allowed to use the \emph{hostname} the software intend to communicate.
However, as long as there is a valid server certificate presented,
even if the software-recognized hostname is not the user's intention, TLS will simply accept it.},
it is now almost useless for protecting casual (careless) users against
phishing.\footnote{If users were careful and paranoid enough so that they check
certificates for every accesses, they can simply protect themselves from
phishing by checking the domain part of the URL to be accessed.}

To overcome the problem of SSL server certificates and phishing attacks, CAs
and browser venders introduced a special kind of SSL server certificates
called ``EV (extended-validation)-SSL certificates''~\cite{caforum-aboutEVSSL}.
CAs has introduced much stricter criteria for issuing EV SSL
certificates, which can almost never be satisfied by phishers.
When web browsers connect to HTTPS servers with EV-SSL certificates
installed, the address bar regions are illuminated green to indicate
that the site is most likely to be a genuine one.
However, as a consequence of the stricter validation criteria,
it can not be introduced for every genuine web sites.
Their intention seems to prevent phishing against some of very high-profile
services such as banking, and not to prevent all phishing attacks.

The opposite to those whitelist approaches are to create and
continuously maintain a up-to-date list of known phishing websites so
that users will not visit those websites.  Most of the today's web
browsers, such as Internet Explorer and Mozilla Firefox has built-in
supports for such ``anti-phishing'' functionalities.  However, such
blacklist can never be complete.  For an instance, according to the documents
which is issued by Microsoft and Mozilla~\cite{mozilla-phishprotect,microsoft2006phishing},
at least 10\% of known phishing sites can not be prevented by those filters.

\subsubsection{Local whitelists}

Another kind of approaches is to maintain a user-specific whitelist
of known genuine hosts by each users.

Petname tool~\cite{close2006petname} is a Mozilla Firefox extension for
providing local whitelist functionality.
It allows users to ``name'' each website they visit.
When the users visit the same site for the second time,
it shows the user-given name with a green background,
showing that the site seems to be the intended one.

Alternatively, most browsers have a built-in support for password managers, which can
be used for this purpose.  When users let browsers to remember a
password for a genuine site, the browsers will automatically
input the password only when the user visits the same site.
The browsers will not input the password for the phishing sites
with a different URI from the genuine sites, thus users can
notice when they visit phishing sites.

Another related approach is to customize login screen for each user by
using HTTP cookies stored in the browser.  An identifier cookie is
stored into each browser, and when a user accesses a login page, the
server customizes the part of the screen using the identifier sent as a
cookie.  As an access to phishing sites does not contain that identifier
cookie, it can only send an uncustomized login screen to the users.

The common problem of those approaches are that these requires local
storages, which hardens users for using several computers
alternately. In addition, users have to maintain the list
appropriately and carefully.

\subsubsection{Existing cryptographic solutions}

There are some cryptographic approaches for hardening web authentication.
SSL client authentication introduces public-key cryptography for
authenticating clients to servers.  Users installs a client certificate
and the corresponding secret key to browsers, and when the server requests
the browser authenticates themselves using the secret key, in the way it is
cryptographically secure against secret stealing and credential forwarding attacks (Patterns II and IV).
The server will be authenticated by using server certificates as usual.
Sometimes it is called a mutual authentication for SSL.  However, in the
context of Web browsers, these two authentication attempts are not tied together:
the server only needs to show any valid SSL server certificate to be
authenticated by the client, and the client will not know whether the client certificate
has been verified by the server or not.  As a result, the client authentication can
not prevent blind acceptance (pattern III) phishing attacks which is
previously mentioned.

Digest web authentication algorithm~\cite{RFC2617} uses MD5 hash
function and a challenge-response style protocol to let clients
authenticated by the servers without sending a plain-text password,
intending to prevent password stealing attacks (Pattern II).  However,
when the client sends the MD5-hashed response to a phishing website, the
phisher can send that value to the genuine site to make a successful
authentication, thus it is vulnerable for credential-forwarding attacks (pattern
IV).  It is also vulnerable for blind acceptance
attacks (pattern III).\footnote {Digest authentication mechanism defines a protocol by
which the server authenticate itself to the client.  However, it is a
optional feature chosen by the server's own, which makes it useless.}
Furthermore, as long as users use passwords with a memorable entropy,
recent computers can easily reverse the hash function by a simple
off-line brute-force dictionary attacks to regain the plain-text
password.  According to a NIST
publication~\cite[Appendix~A]{NIST-SP800-63}, user-inputted passwords of
40 characters only have about 62 bits of entropy, which is considered
insecure against off-line attacks nowadays.  It means that the Digest
authentication can not actually prevent plain-text password to be
obtained by phishers and eavesdroppers.

PwdHash~\cite{ross05PwdHash} overcome the credential-forwarding attacks by
generating a distinct hashed temporary password for each websites
from a single passphrase.  When user input a passphrase preceded
by two \texttt{@} characters to any password entry form field, PwdHash
traps the keystrokes and inserts a temporary password generated by
hashing the input passphrase with the hostname of the form's target.
When user unwittingly inputted a passphrase to a phishing site
(with \texttt{@@}-prefix), the generated temporary password will be different
from the one valid for the genuine site, thus preventing 
password stealing (pattern II) and forwarding (patterns IV \& V) attacks.

However, PwdHash can not provide mutual authentication, thus it can not
prevent blindly-accepting (Pattern III) attacks. In addition, the off-line attack
can regain the user's password in the same way as for the Digest authentication,
which makes it ineffective against attacks in pattern II.

\section{Our proposal}\label{sec:proposal}

\subsection{Design goals and criteria}

Considering the weakness of existing countermeasures against phishing,
we designed a new authentication protocol which can easily prevent all
kinds of phishing attacks. The design of the protocol considers
future replacement of HTML form-based authentication mechanisms implemented in Web-application layer
as well as Basic and Digest authentication mechanisms
defined in \cite{RFC2617}. To realize this
goal, the design must be carefully considered so that it does not make
existing web applications difficult to adopt the new protocol.

While designing the protocol, we have settled several goals and criteria for it.
\begin{description}
 \item[Password-based authentication without local storage]
	    Novice users are well accustomed to the password-based
	    authentication, and it is difficult for them to manage local
	    secrets (such as private keys) stored in the
	    local hard-disk in the correct way.  In addition, the
	    existence of local secret also makes difficult for users to
	    use more than one computer simultaneously.  The protocol avoid
	    requirements for any long-term local secret storages and/or local whitelists.

 \item[Secure mutual authentication] Of course, the protocol must be
	    secure enough to prevent phishing attacks.  Especially, we
	    focus on that it provides mutual authentication to prevent
	    blind acceptance attacks (Pattern III), which was not achieved
	    in the above-presented existing countermeasures.
	    Of course, it should be secure against other types of
	    attacks including forwarding attacks (Patterns IV and V).

	    The design also considers that the protocol should secure against
	    off-line attacks even after the client has been started
	    communication with phishing websites, because normal users
	    is unlikely to be able to set up a password which is
	    resistant to off-line dictionary attack.

	    There is an Internet-draft proposing introduction of Mutual authentication for preventing
	    phishing attacks as mentioned in~\cite{hartman-phishing}, but
	    none of proposed/widely-accepted countermeasures provide this feature for HTTP.

 \item[Generic (equal opportunity)] The protocol should be usable to \emph{any} web services who
	    want to provide user authentication, including personal
	    websites.  Current EV-SSL scheme does not provide this
	    criteria: EV-SSL certificates can only be acquired by
	    parties having long-enough credit as a commercial entity.

 \item[Compatibility with real Web applications and systems]
	    The protocol should be designed so that it can be easily
	    deployed for many real Web application with complex needs
	    and heavy loads.  Many existing solutions, including Digest
	    authentication fails in this criteria.

	    In more detail, we focused on the following topics.

	    \begin{description}
	     \item[Per-URL authentication configuration,
			Multiple authentication realms on one server] 
			In real web systems, authentication is defined for sets of
			web resources specified by URIs.  For example, most Web applications
			have both public contents (e.g.\ blog contents) and private, 
			authenticated contents (e.g.\ blog editing interfaces).
			The protocol should be able to set up authentication only for
			the part of the resources on one Web server, unlike TLS client authentication.

			In addition to the above, there are web applications which have
			multiple set of authentication ``realms'' (groups of resources which
			share same setting of the authentication) on one server.

			For example, the mail interface of the ``Google Apps''\footnote{\texttt{http://www.google.com/a/}}
			sets up an independent authentication realm for each domains hosted:
			The main Gmail service has the URI \url{http://mail.google.com/},
			while the hosted service for example.com domain has the URI
			\url{http://mail.google.com/a/example.com/}.  Other examples are
			the customization interface for hosted mailing lists, a hosted
			environment for wikis,~etc.  When multiple realms are not supported,
			such services must set up individual host names for every realms,
			which is so cumbersome that unlikely to be deployed.
	    
	     \item[Sufficient support for recent Web application designs]
			One strong motivation for
			current web application designers/implementors to use
			form-based authentication instead of HTTP Basic
			authentication is the flexibility of authorization.
			When authentication is implemented using forms and HTTP
			cookies, the Web application can control various aspect of
			authentication/authorization.  For example, such applications can provide
			both unauthenticated (guest) contents and user-customized
			contents on the same URI depending on the status of
			authentication, set timeout for user's inactivity or total
			login time and force logout, and implement explicit
			``logout'' UI to forget authentication status so that the
			same terminal can be used for multiple user identities
			(including unauthenticated guests).

			However, form-based authentication is inherently vulnerable
			for phishing attacks, because the behavior of the forms is
			fully controlled by the web contents.  To the purpose of the
			security, we have to avoid use of form authentication as
			well as existing HTTP authentication.
			To replace form-based authentication with our new protocol,
			the protocol provides several functionalities for 
			implementing such applications on the top of
			this protocol.

	     \item[Compatibility with load-reducing equipments] Many web
			services with heavy loads uses several existing
			equipments for reducing the server loads.  For
			example, many services uses SSL accelerators,
			which are reverse-proxies dedicated for talking
			TLS protocol and forward traffics to the
			back-end servers.  They also uses a
			load-balancing reverse-proxies with multiple
			back-end servers for the same contents.

			We designed the protocol so that most of the
			these facilities can be used with our
			authentication protocol as much as possible.
			Some example configuration for deploying our
			protocol in large systems are shown in Section~\ref{sec:deployments}.
\end{description}
\end{description}

\subsection{Designing Policies}

Given the above criteria, we have designed the protocol having the
following properties.

\subsubsection{Use of password authenticated key exchange (PAKE)}

To prevent password to be revealed to the phishing sites even with
off-line attacks, we have chosen a variant of PAKE (Password-based
Authentication and Key Exchange) protocols as an underlying
cryptographic protocol.  PAKE protocols enable mutual authentication
between clients and servers by using only a shared ``weak secret'' (such
as passwords), while keeping weak secrets not known to the both party on
the failed authentication.  By this property, when users have been
connected to the wrong server (possibly a phishing server), and when
they try to authenticate themselves using this protocol, the users'
passwords are kept secret to the server, unlike Basic/form
authentication (passwords are sent in plaintext) or Digest
authentication (vulnerable against simple blute-force attacks)
mechanisms.  This prevents password stealing attacks (Pattern II)
completely.  Furthermore, the protocol benefits from PAKE to detect
blind acceptance attacks (Pattern III).  Mutual authentication provided
by PAKE protocols will never succeed when the server does not know about
user's password.  The protocol, unlike previous ones, checks successful
authentication on both client and server side to detect and prevent such attacks.

\subsubsection{Detection of credential forwarding attacks}\label{sec:credforward}

PAKE is actually a key-exchange protocol which is usually used for both authentication and encryption.
Under such use case a man-in-the-middle attack (MITM) is impossible, because the encryption key established
by the key exchange is known only to the both peers of the PAKE negotiation.
However, because the protocol implements authentication in the
HTTP communication layer, we use established secret shared key only for the authentication.
In this setting, PAKE itself can not provide security against MITM, thus
we need another mechanism to prevent credential forwarding attacks.

Our solution is to modify PAKE protocol to check that intended host-name is
matched at both authenticating endpoints.  This check is sufficient for
preventing phishing, because forwarding phishing sites uses hostnames which is
different from the genuine servers.  When a client talks to a phishing websites
which forwards the received messages to another website, the
authentication will fail because the weak secret and/or confirmation
materials does not match, preventing pattern IV (and V) attacks.
The secret password will not be revealed to the phishing servers also in this
case, by the virtue of the behavior of PAKE protocol in secret mismatch case.

\subsubsection{Extending HTTP-layer authentication architecture}

To achieve authentication handling depending on the requests, we have
designed the whole protocol as an extension to the
current HTTP protocol~\cite{RFC2616} and HTTP authentication
mechanisms~\cite{RFC2617} in the HTTP message layer.  The protocol is
naturally designed using existing architectures of HTTP authentication
as far as possible, so that it can be easily integrated to existing HTTP
implementations.

One problem of authentication tied to the connection layer, such as TLS
client authentication, is that the authentication runs before the client
send the request to the server, therefore server can not know what
resource the client is accessing to.  There is also a gap between 
connection layer authentication and the HTTP keep-alive architecture.
The proposed protocol avoids such deficiencies.

On the other hand, there is several weaknesses on applicability of both
HTTP-based and connection-based authentication for web applications as
described above.  We are solving those weaknesses by slightly extending
the HTTP authentication architecture (see
Section~\ref{sec:morefeatures}).

\section{The protocol}\label{sec:protocol}

This section describes the overview of the proposed protocol in this
section.  The authentication is based on a PAKE variant called the ``Key
Agreement Method 3 (KAM3)'', originally proposed by Kwon, defined in the
ISO standard ISO-10770-4:2006~\cite{ISO-10770-4}.  In this section, only
a typical conversation for the protocol using discrete-logarithm-based
settings is described: the full specification is available as an IETF
Internet Draft~\cite{draft-oiwa-http-mutualauth-05}.

An example session trace for the whole process of the protocol
negotiation will be shown in Appendix~\ref{apd:log}.

\subsubsection*{Notations}

In this section, we let $q$ be a prime number defining a finite group, 
$g$ be a generator inside the ``$\bmod q$'' group,
and $r$ be a rank of the subgroup generated by $g$.
The generator $g$ must be carefully chosen so that $r$
becomes a large-enough prime (in the specification we set the
parameters to satisfy $r = (q - 1)/2$).
$H(\tilde x)$ is a hash value generated from elements of list $\tilde x$.

\subsection{Preprocessing (registration phase)}

Before authentication starts, there are some things to be set up.
The authentication is performed using a username and a secret password.
A pair of a username and a password is valid for
one ``authentication realm'', a group of Web pages on which 
the same set of users and passwords are valid.

In the protocol, each authentication realm is defined as follows:
a parameter ``auth-domain'' specifies
the range of the hosts using the same set of usernames and passwords (authentication domain).
It can be a single host (FQDN, e.g. ``\url{www.example.com}''), 
or it can be limited a specific protocol and port of the host
(e.g. ``\url{https://www.example.com:443}'').
It also supports a realm which is a set of hosts in the same
domain (e.g.,~``\url{*.example.com}'') to support single-sign-on services.
Another parameter ``realm'' is a simple string which is used as a label
inside each authentication domain.
Each pair of auth-domain and the realm defines one authentication realm.

For each authentication realm, a set of valid username-password pairs is
defined.  The user has one pair of the username and the password in plaintext.
The weak secret $\pi$ used by the client for PAKE-based authentication
is extended from the ISO specification to include
the authentication domain, the realm and the user-name, defined as
\[
 \pi = H(\text{algorithm}, \text{auth-domain}, \text{realm}, \text{username}, \text{password}).
\]
The server will only need its specially encrypted version $J(\pi)$ defined as
\[
 J(\pi) = g^\pi \bmod q.
\]
The retrieval of the password from $J(\pi)$ needs brute-force searching.

\subsection{Message Exchanges}

For the first access to the authenticated contents,
the protocol requires three round-trips of HTTP messages.
First, as a response to a client's request (without any
authentication credentials), the server sends a usual HTTP 401
response to request authentication.  This is very similar to
the Basic and Digest HTTP authentication.
The response from the server specifies an ``algorithm'' which defines
the parameters $(g, p, r)$, an authentication realm and a realm to be
used.

The client asks the user for the username and password.
When the password is available from a user input, the client
constructs a cryptographic value $w_a$, which is generated from a random
number $s_a$ as
\[
 w_a = g^{s_a} \bmod q,
\]
and sends the second HTTP request along with the user-name.

The intermediate 401 response from the server 
contains another cryptographic value $w_b$, which is generated from
$J(\pi)$ stored in the server-side password database and another random number
($s_b$) as
\[
  w_b = (J(\pi) \times w_a^{H(1, w_a)})^{s_b} \bmod q.
\]
Because this value have randomness from $s_b$ which is larger than the
entropy of $J(\pi)$, $w_b$ can not be distinguished from a random number
by any other party. This means that the value of $J(\pi)$ can not be
extracted from $w_b$ in any way unless one knows $\pi$.  The equations
for $w_a$ and $w_b$, are directly derived from the ISO specification.

At this time, both the client and the server compute a shared secret $z$
using different set of known values.  The client calculates it from
$s_a$, $w_a$, $w_b$ and $\pi$ as
\[
 z = w_b^{(s_a + H(2, w_a, w_b)) / (s_a * H(1, w_a) + \pi) \bmod r} \bmod q,
\]
and the server calculates it from
$w_a$, $s_b$, $w_b$ as
\[
 z = (w_a \times g^{H(2, w_a, w_b)})^{s_b} \bmod q.
\]
If and only if a correct pair of $\pi$ and $J(\pi)$ is used, the two $z$
calculated at both sides will match.  As long as eavesdroppers do not know either $s_a$
or $s_b$, $z$ can not be reconstructed because this key exchange is a modification of
Diffie-Hellman key exchange protocol.

The final step of the protocol is to check whether mutual authentication is
succeeded by confirming the value of $z$ at each side.
the client sends a third request containing value of $o_a$.
The value $o_a$ is a hash value defined as
\[
 o_a = H(4, w_a, w_b, z, nc, v),
\]
where $v$ is the host verification element defined in Section~\ref{sec:verification},
and $nc$ is the value of nonce counter.
The server calculates the same value by its own value of $z$ and verifies
the equality.  If they matches, the server grants the access to the resource
and sends the final response with the value $o_b$, which is generated in the similar way as
\[
 o_b = H(3, w_a, w_b, z, nc, v).
\]
The receiving client \emph{must\/} verify the value $o_b$ to check
whether the server is genuine.  Without knowing the true value of
$J(\pi)$, the server can not construct the correct value of $o_b$.
This means that phishers can not forge clients by blindly accepting
authentication requests, thus Pattern III attacks are prevented.
The equations for $o_a$ and $o_b$ are also extended from the ISO 
specification to include various additional values to be verified,
such as $nc$ and $v$.

When a second request to the same host is sent, the client can reuse the
shared secret $z$ by directly sending the third message to the server using
the same session identifier and $z$, incrementing the nonce counter.
In this case, both clients and servers only need to perform a
lightweight hash operation but no public-key operations.

\subsection{Host verification element}\label{sec:verification}

Host verification element $v$ ensures that the client is directly
communicating to the server on which the genuine web service intended to
do so\footnote{the protocol allows services
to safely introduce a relaying facility on their intention, while preventing
unwanted relaying attacks.  See Section \ref{sec:deployments} for some examples.},
and no credential forwarding attacks are employed.
The value to be used for such verification must have the following properties: (1) it is
shared between the client and the server, and (2) when talked to
the phishing site, it will be different from what will be used for the
genuine site. This protocol uses the following values, chosen by the
``validation'' field in the response header, for that purpose.

\begin{itemize}
 \item ``\texttt{Host}'' verification: the string $v$ will specify
       the scheme-host-port triple, for example, \hfil\break``\url{http://www.example.com:80}''.
 \item ``\texttt{Tls-cert}'' verification: $v$ will be the hash of the
       server public-key certificate which the HTTPS server uses.
 \item ``\texttt{Tls-key}'' verification: $v$ will be the master secret of
       the underlying TLS sessions.
\end{itemize}

For HTTP services the Host verification is used, as in this case
transport-level security is not intended.  For HTTPS services,
we can use any of the above three methods safely, although we recommend
\texttt{tls-cert} mainly for applicability.
See Section~\ref{sec:TLS-verify} for some discussions.

\section{User Interface}\label{sec:ui}

A possible attack to this protocol is to forge the user interface
asking a password.
Existing Basic and Digest authentication usually uses a modal dialog box
for asking usernames and passwords.  However, it can be easily imitated
by using either a stylized HTML, an HTML pop-up window or even a picture of
a pop-up dialog box.  In general, any region inside the content area is always
vulnerable for forging by phishers.  In this context, the weakness of the forms inside
the document is needless to say.  If plaintext passwords are
sent to the phishers outside of this proposed protocol, we can not prevent 
any patterns of phishing attacks.

To prevent such kind of attacks, we propose to use the ``chrome area''
(the area where an address bar and other browser UI components are
exist) for the new authentication protocol.\footnote{It is extremely
important that the proposed UI is only used for this protocol (and more
secure protocols, if any) and not for weaker authentication mechanisms
such as Basic and Digest authentications, so that to it is ensured that
the password entered to this UI is not sent in an insecure way.}  Our
implementation of the Mozilla Firefox extension uses address-bar area
(where web-pages do not have any access to) for password input instead
of using pop-up dialog windows.  Note that the chrome area is considered
to be a secure region, and if there were a method to overwrite there, it
shall be considered to be a vulnerability of the
browsers~\cite{microsoft2004-ie-winrestriction}\footnote{In the past,
there was such implementation problems to allow forgery of browser
security indications, but now all known problems are considered as
security vulnerabilities and fixed.}.  Inside the address bar we put
input boxes for a username and a password, and also an indicator for
displaying the authentication status.

\begin{figure}[t]
 \begin{center}
  \noindent\includegraphics[width=0.6\textwidth]{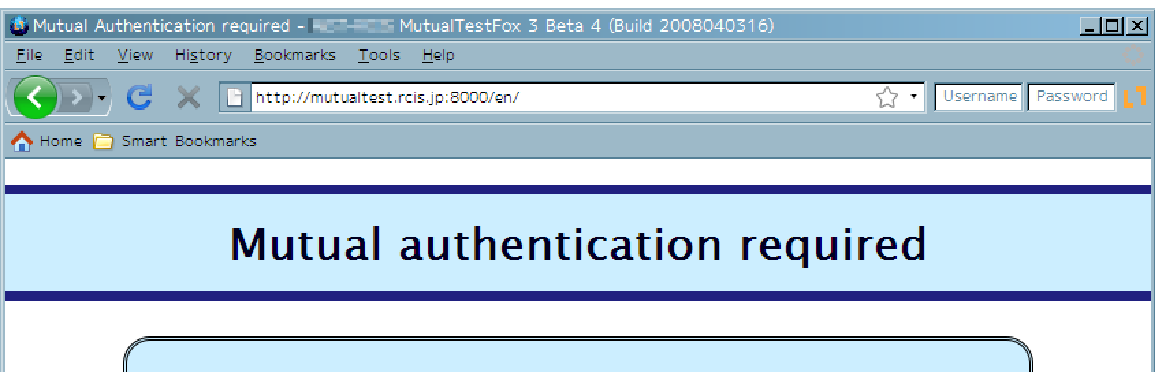}\\
  (a) Mutual authentication is requested.\\
  Input fields are in the chrome area beside the address bar field.\\[2\baselineskip]
  \noindent\includegraphics[width=0.6\textwidth]{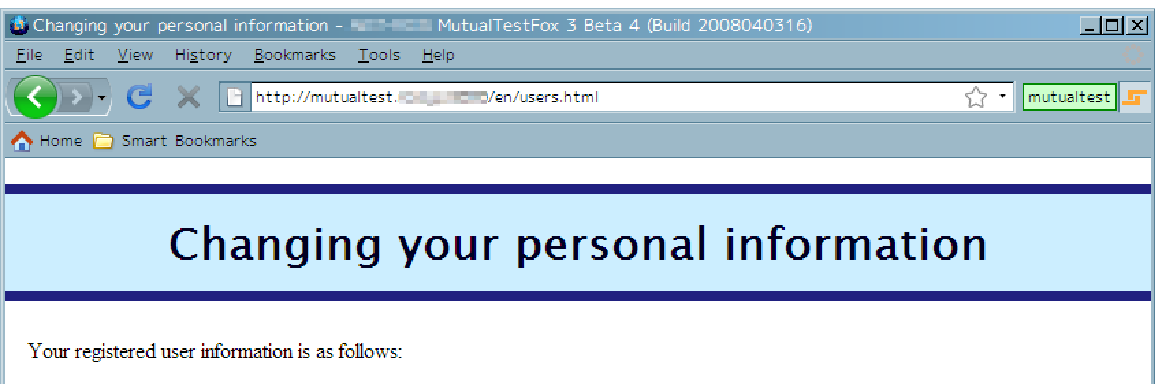}\\
  (b) Mutual authentication has been succeeded. \\
  Username ``mutualtest'' is displayed in the chrome area.\\
\begin{footnotesize}
\end{footnotesize}
 \end{center}
\caption{Chrome-area input fields for user-names and passwords}\label{fig:UI}
\end{figure}

In our implementation, the input boxes are initially not displayed.
When users visit a website which requires mutual authentication,
two input boxes for a username and a password automatically
appear on the right of the address bar (Figure~\ref{fig:UI}~(a)).
When the user enters the user-name and
the password the authentication begins, and after the authentication success
has been recognized by the client (by checking the final $o_b$ value
in the response message), it displays a mutually-authenticated user-name
with a green background on the right of the address bar (Figure~\ref{fig:UI}~(b)),
showing that the identity of the host is verified by the password.

By using this protocol and the UI, users can prevent all patterns of
phishing attacks which involve passwords (see patterns II--V in
Section~\ref{sec:phishing}) by following two simple rules:
\begin{itemize}
 \item Users must enter password only to the text-box provided
       by this UI.
 \item Users must enter private information such as credit-card
       numbers only when the indicator becomes green.
\end{itemize}
If a user sees the green indicator, user can trust that the
communicating service is the genuine one (whom the user has told
his/her password before).\footnote{Of course, it \emph{does not} mean
that the peer will use the credit-card numbers righteously.  If users use
``untrustful'' services as their intention, it is up to their
responsibility.  Our protocol's responsibility is to prevent
communicating with untrustful services without the users' intention.}

Furthermore, if this protocol is widely accepted and introduced to 
many commercial websites, we can expect that the above rules prevents even
phishing attacks without passwords involved (Pattern I attacks), because
when users are well trained to always use Mutual authentication, they
will find that such phishing sites are requiring an unsafe
authentication which is not the usual.

The only exception to the above rules is the initial enrollment to the web
services.  In the current design, users must input a plaintext password
to somewhere in which it is sent in raw password (possibly) over the
TLS secured channel. Future version of the protocol may support sending encrypted
password token $J(\pi)$ to the server, but it does not completely solve
the problems in such situation.
In the time of initial enrollment, users must always be
very careful to determine whether they should continue to access the web
services, because they need not only to check the connection integrity,
but also to check the service's social trustworthiness which is 
completely outside of the scope of authentication protocols.

\section{Additional features for Web applications}\label{sec:morefeatures}

We have designed the protocol as it is almost upper-compatible with
existing HTTP authentication mechanism.  If websites are
using Basic or Digest authentication, it is easy to migrate to the
new protocol by reconstructing the password database
from plaintext passwords.\footnote{For Digest authentication,
the protocol even provides a backward compatibility feature so that
the database can be directly converted for the new mutual-authentication protocol.}

However, most existing websites are using form-based password
authentication implemented in web application level instead of HTTP-layer
authentication.  There may be many reasons to do so including
UI design flexibility, but we think that the strongest reason
is the lack of flexibility in HTTP Basic and Digest authentication.
We have extended a base mechanism of HTTP authentication in several ways
to overcome these problems with the mutual authentication protocol.

\subsection{Optional authentication}

Today's typical web services often accept guest users and
authenticated users simultaneously.  Such a service shows a default
contents for the guests when a user accesses it for the first time, and
when the user authenticates themselves it shows a content customized
for each user.  For example, in 
a commercial auction website, summaries
of exhibited items are displayed for all users including unauthenticated guests, but
a form asking a bid price is displayed only for authenticated users.  It is
easy to implement this kind of services by using form-based
authentication, but it is tricky to implement with HTTP Basic/Digest
authentication.

\begin{sloppy}
The specification of the new protocol introduces a direct support for such services.
Web servers can ask clients to \emph{optionally} start mutual
authentication while sending contents for guest users by
using a newly-introduced ``\url{Optional-WWW-Authenticate}'' header.

In websites using optional authentications, Web servers will send normal
``200 OK'' responses with a newly-defined
``\texttt{Optional-WWW-Authenticate}'' header.  The browser will render
the response body (containing a page for guest users) in a usual way,
and at the same time it enables the input fields for authentications.
If users wish to visit the website as a guest user, they can
just continue browsing as usual.  On the contrary, if they want to use
website as authenticated users, the browser will simply send a
Mutual authentication request by the users' requests, and servers will
respond as usual.
Our new UI design for this authentication protocol (Section~\ref{sec:ui})
is well suited for this use, because it is implemented as non-modal
input fields.
\end{sloppy}

Implementation for this additional feature will not be difficult:
In the server side, semantics for a ``200 OK'' response with an
\url{Optional-WWW-Authenticate} is almost similar
to a ``401 Authentication Required'' response with a \url{Optional-WWW-Authenticate} header.
Also in the client side, the effort for implementing optional
authentication will be almost as same as that for implementing
the Mutual Authentication itself.  Our implementations already support
this in both sides.

\subsection{Application-level control for login/logout behavior}

Many web applications have their own policies for handling the
authentications.  For example, many applications implements a forced
time-out for login sessions for either user's inactivity time or the
time from the beginning.  In some cases applications want to forcibly
terminate an user's authentication sessions.  Some application forces
entry-point of the authenticated pages to on location.

However, such control was one of the weak points of existing HTTP
authentication mechanisms.  In current HTTP authentication mechanism,
once a user inputs the user-name and the password to a browser,
it continues to send them to the website until the user closes all
browser windows, and there was no way for servers to specify that
the current log-in session is to be terminated, except for requesting
a new credential by sending a false ``login failed'' response.
In addition, users are also difficult to switch to another account
without closing the browser once.

The proposed protocol introduces a new header for a finer control of the
browser's behavior related to an authentication.  Using the new header,
named \texttt{Authentication Control}, servers can specify that the
client should forget login passwords after a given time period.  It also
allows servers to specify a conditional redirects to another URL based
on a current status of authentications, e.g. a redirect only for a user
not starting an authentication.  By using this header, applications can
use many of current designs of webpage structures (such as fixed log-in
page or log-out page) as is, along with our new authentication
mechanism with little modifications.
Our UI implementation also allows users to logout the service by their
own.

\subsection{Domain-level single sign-on (SSO) support}

Recent large-scale websites such as Yahoo!\ or Google use several hosts
for serving a group of services.  In the HTTP Basic authentication,
however, the realm of the authentication (the area that shares the same
user-name and the password) is defined by the pair of the host-name and
the realm value specified in the response header, which means that the
same password must be reentered for every host even in the same domain.

The current version of the protocol supports this kind of websites by allowing a wild-card
for the authentication domain.  If the authentication domain is set to
``\texttt{*.example.com}'', for example, the same user-name and password
will be automatically used for any host inside the domain
\texttt{example.com} when the same realm value is used.  While passwords
are shared, each servers inside the authentication domain will need an
access to valid password databases, because the host verification
element $v$ still differs between servers.  This forbids credential
forwarding even within the same authentication domain.  As a result,
phishing will be impossible even when there is a less-trusted host
inside the domain as long as the password database is kept
secret.

\section{Implementation}\label{sec:implementation}

We already have several implementations of the protocol, most of which
are published as open-source software.  These software can be downloaded
from our project homepage\footnote{Project Homepage URL: \url{https://www.rcis.aist.go.jp/special/MutualAuth/}}.

For the server side, we implemented an extension module for the Apache
2.2 Web server.  The module, \texttt{mod\_auth\_mutual}, supports basic
functionality of the Mutual authentication protocol.  It can be used
almost in the same way as existing authentication modules such as
\verb|mod_auth_basic| and \verb|mod_auth_digest|.  We have also
implemented a preliminary module for the WEBRick server written in the
Ruby language.

For client side, We have two implementations: our \emph{MutualTestFox}
(shown in Figure~\ref{fig:UI}), which is a modified version of
open-source Mozilla Firefox, implements both back-end protocol and
proposed UI design.  We also have an Internet Explorer-based
implementation based on Lunascape\footnote{\url{http://www.lunascape.jp/}}, a proprietary browser implementation.

There are two experiment Web sites which are open to public.
Especially, Yahoo!\ Japan Auction have deployed in the last year a
special experimental trial site on which users can log onto the trial
site using their existing Yahoo!\ ID and passwords, using both client
implementations.

\subsection{Performance}

We designed the protocol carefully so that the computation resources
required for server side will be as low as possible.  If an
authentication session takes its first place, there will be three HTTP
round-trips (request-response pairs).  There is one round-trips more
than the Basic or Digest authentication methods, but unlike Digest
method, there is no need to generate any random numbers until the client
really need to authenticate itself to the server.%
\footnote{HTTP Digest authentication needs random nonces
for every responses from servers, even if the client do not want to
start authentication.}  There will be two computations similar to RSA
secret-key operation (a power-modulus with large exponents), and two
cryptographic-hash computations.  This overhead is similar to the SSL
client authentications.

Once an authentication has succeeded, a ``cryptographic session''
associated with a cryptographic secret is shared between a server and a
client.  For the second (or later) request, authentication can be done
in a very light-method way with only one hash operation (no public-key
cryptography operations).  These sessions are designed to be implemented
in a constant memory (including nonce duplication checks), and can be
discarded in any time under the server's decisions.

On the real web server we're currently using (a server with 1.73GHz
Intel Pentium M processor and 512MB memory), an authentication request
without pre-shared secrets takes only about 132 milliseconds for
processing (using 2048-bit discrete-logarithm and SHA-256 setting)\footnote{We measured a time consumed in our authentication
handlers inside Apache web server, without network and other overheads.}.
After a secret has been shared between peers, processing the second
request only requires 0.4 milliseconds.

\section{Discussions}\label{sec:discussion}

\subsection{Choice of host verification elements}\label{sec:TLS-verify}

As described in Section~\ref{sec:verification}, we provide three
possible values used for host verifications.  Strictly speaking, the strength
of the verification is \texttt{tls-key} the strongest, \texttt{tls-cert}
the next, and \texttt{host} the weakest.  \texttt{Tls-cert} verification
assumes the secrecy of the secret key associated with the certificate,
and \texttt{host} verification assumes correct operation of the CA and
PKI even for applications outside the Web.

The reason for us to provide the solutions other than \texttt{tls-key}
is the difficulty of the implementation of the \texttt{tls-key}
verification.  If every details of the TLS protocol implementation is
available, it is possible to implement the strongest
\texttt{tls-key} method. However, for the most black-box implementation of TLS
protocols, it is not possible to acquire the shared secret key of the
underlying connections.  \texttt{Tls-cert}, on the other hand, works
quite well with black-box TLS implementations because it only uses a public
information.  It also works well with environments using TLS
accelerators, important for large-scale installations, without
sacrificing security.

\subsection{Possible deployments}\label{sec:deployments}

The protocol design enables several flexible ways of deployment for existing web system
with minimal modifications.

\begin{figure}[t]
\center{\noindent\includegraphics[width=0.6\textwidth]{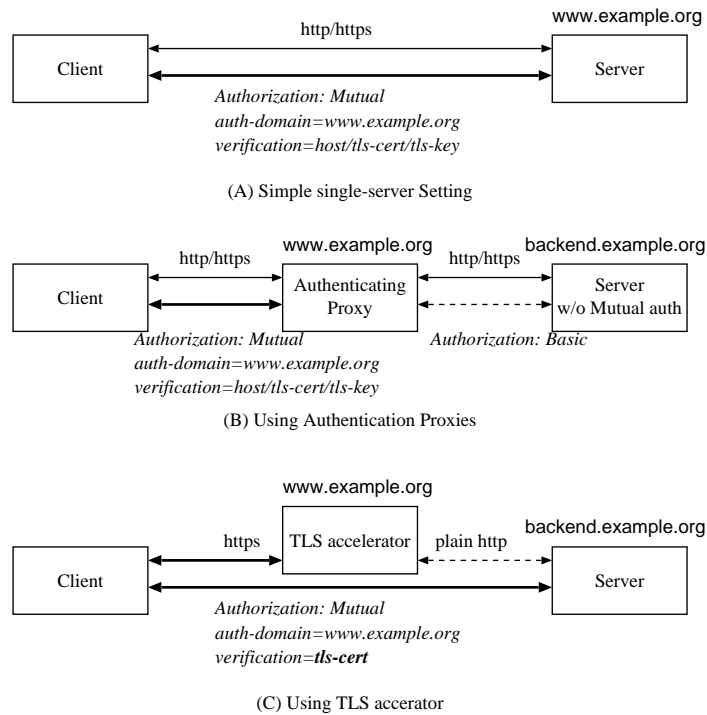}}
\caption{Possible deployment for Mutual authentication}\label{fig:deploy}
\end{figure}

\begin{enumerate}
 \item Single-server setting

       For the small services, the server-side authenticator can simply
       be installed in the web server (Figure~\ref{fig:deploy}-A).
       We have implemented an extension module for the Apache web
       server, which can be used almost in the same way as other
       authentication modules such as mod\_auth\_basic.

       The simpler one of our experiment websites uses this setting.

 \item Authenticating reverse-proxy

       If there are web servers which are difficult to modify its
       setting, we can implement authentication in reverse proxies
       (Figure~\ref{fig:deploy}-B).  The reverse-proxy selectively adds
       mutual authentication requests into the responses from the
       back-end server, and when authentication has been succeed,
       the proxy passes the authentication results to the back-end
       server using additional headers.  This reverse proxy can also work
       as a load-balancer for several back-end servers.

       Our large-scale experiment on the Yahoo!\ Auction trial site uses
       this scheme, since the existing service servers are not possible
       to implement Mutual authentication directly.

 \item TLS accelerators

       Many existing web services uses dedicated, single-function 
       SSL accelerators to reduce SSL encryption overhead on the application servers.
       The proposed protocol can cope with such systems  (Figure~\ref{fig:deploy}-C).
       When TLS accelerators are used, 
       TLS encryption negotiations are done between clients and
       the accelerator, and the requests sent encrypted from the client
       are forwarded to the back-end web server in plain-text form.
       In this case, it is impossible to perform user authentication in the TLS-level,
       because such accelerators does not have anything but simple TLS negotiation functions.
       However, our protocol works under such deployment,
       as long as the back-end server knows that the accelerators are used and
       the \texttt{tls-cert} host verification is used.
       The back-end server speaks plaintext HTTP protocol and
       negotiates the Mutual authentication as if it is using the HTTPS protocol.
       As long as the accelerators forward only responses from the genuine servers,
       phishing attacks are correctly prevented.
\end{enumerate}

\section{Related work}\label{sec:relatedwork}

Several related works are presented in Section~\ref{sec:countermeasures}.
In this section we describe about TLS-SRP, recently proposed extension to TLS.

TLS-SRP~\cite{RFC5054} is an extension to the TLS encryption protocol
which uses SRP, another variant of PAKE protocols, both for authentication
and key exchange.
TLS-SRP can be used either with or without a server certificate issued
by PKI certificate providers, and performs SRP key exchange during the
negotiation phase.
It may be suited with several TLS applications which basically use one
connection per client, such as IMAPS, POPS, and TLS-based VPNs,
because all of these protocols perform authentication and application-specific 
operations sequentially on one TCP/TLS connection, so that TLS-SRP 
can naturally replace the first authentication phase.

While we designed the proposed protocol, we have examined the use of
TLS-SRP for preventing phishing against Web application systems
along with the current HTTPS protocol, and we have concluded that
it is not well suited for Web uses for several reasons.
While TLS-SRP's authentication is strongly bound to the underlying TCP connection, 
HTTP transport is not used in such way:
in HTTP, concurrent HTTP requests to several applications in
the same web server are sent sharing one or several TCP or TLS connections
intermixed.  Furthermore, the keep-alive feature implemented in HTTP/1.1
allows several requests to hosts sharing the same IP address to be streamlined in
one underlying connection, either
TCP or TLS.  When TLS-SRP is used for Web application authentication,
two requests for different web applications
with separate authentication and authorization may be sent on the
same TLS-SRP connection authenticated with a single identity, which
is conceptually undesirable.

Moreover, because TLS-SRP negotiation is
performed before establishment of TLS encrypted channel (i.e.\ before
the detail of requests are sent), there is no way
to negotiate any detail of the authentication (such as realms)
before investigating the request.
It forces all contents on one server to share one authentication realm,
which sometimes requires large-scale redesign of the application.
For those reasons, we abandoned to use
TLS-SRP and have implemented PAKE-based authentication inside the 
HTTP protocol.

\section{Conclusion and future directions}\label{sec:conclusion}

We have designed a new secure password-based Web mutual authentication protocol
which prevents various kinds of phishing attacks.  The protocol design considers
various aspects of existing web systems, such as additional requirements from
applications, systems with multiple hosts, and various ways of system deployments.
We also propose new browser UI design to securely use this protocol.

Our implementation of the browser extension and
the web server module are available as
open-source software.  We are also performing
demonstration and experiments in a part of the `Yahoo!\ Auction''
website, which is the largest auction website in Japan.
During the experiment we will examine applicability of this protocol
to the existing Web applications, examine the user experiences,
and use these results to improve the protocol
specification for standardization.

We currently consider to extend the protocol to support various
existing authentication applications.  For example, supporting secure
one-time password tokens and inter-domain single sign-on protocols may
be desirable.
We hope this protocol to be
widely-accepted and deployed in many websites to defeat phishing attacks.

\bibliographystyle{plain}
\bibliography{mutual-auth}

\appendix

\section{An example session log}\label{apd:log}

This appendix shows an example log for a session performing a Mutual
authentication.  Large cryptographic values and other undetermined
values unrelated to the protocol is replaced to placeholders like \texttt{...}.

Like existing HTTP authentication, the client first send a request
without any authorization credentials, and server requests an
authentication in a 401 response.

\begin{alltt}\small
GET / HTTP/1.1
Host: www.example.com

HTTP/1.1 401 Authentication required
WWW-Authenticate: Mutual version=-draft05, algorithm=iso11770-4-dl-2048, 
        validation=host, auth-domain="www.example.com",
        realm="Protected Contents", stale=0
Content-Type: text/html; charset="ISO-8859-1"
Content-Length: ...
\end{alltt}

In the second round-trip messages, the peers performs a password-based
key-exchange.

\begin{alltt}\small
GET / HTTP/1.1
Host: www.example.com
Authorization: Mutual version=-draft05, algorithm=iso11770-4-dl-2048, 
      validation=host, auth-domain="www.example.com", 
      user="foobar", wa="0EeAHWPU4Izqrag4vuMs036...VhGBTrdS0YUdlAE7+J2=="

HTTP/1.1 401 Authentication required
WWW-Authenticate: Mutual version=-draft05, sid=d9ea626480044abd, 
      wb="RUF+vO7/uSQ+/t+uzsV3mkL5/6...TE2w+9HyB6c88+Npptedy==",
      nc-max=256, nc-window=64, time=300, path="/"
Content-Length: 0
\end{alltt}

The third, final exchange for the first request confirms whether the
authentication has been succeeded.  Not only the server, but also the
client confirms the mutual authentication using the value \texttt{ob}
returned from the server in the \texttt{Authentication-Info} header.
The Authentication-Control header specifies the browser should 
terminate the log-in session 300 seconds later.

\begin{alltt}\small
GET / HTTP/1.1
Host: www.example.com
Authorization: Mutual version=-draft05, algorithm=iso11770-4-dl-2048, 
      validation=host, auth-domain="www.example.com", 
      user="foobar", sid=d9ea626480044abd, nc=0,
      oa="33aVf+9Vgtdjh7S...S6NmleE/IFy="

HTTP/1.1 200 OK
Authentication-Info: Mutual version=-draft05,
      sid=d9ea626480044abd, ob="K6FkRV4gFy+XLh...Ow9gAAVhYkSg="
Authentication-Control: logout-timeout=300
Content-Type: text/html; charset="ISO-8859-1"
Content-Length: ...
\end{alltt}

Once a key-exchange has been succeeded, the shared key is
reused for several HTTP requests within the browsing session.
The client uses the same session ID (sid) with an incremented
nonce count (nc) to request the authentication using the previously-shared secret.
The Authentication-Control header in the response reset a log-out timer
set by the previous response to 300 seconds again.

\begin{alltt}\small
GET /page2.html HTTP/1.1
Host: www.example.com
Authorization: Mutual version=-draft05,
      algorithm=iso11770-4-dl-2048, 
      validation=host, auth-domain="www.example.com", 
      user="foobar", \textbf{sid=d9ea626480044abd, nc=1},
      oa="U6wm8+IlkhNmdM...33x0/wnLfz="

HTTP/1.1 200 OK
Authentication-Info: Mutual version=-draft05,
      sid=d9ea626480044abd, ob="0VV8C+KZsT6+...rN2vbKANiDoez="
Authentication-Control: logout-timeout=300
Content-Type: text/html
Content-Length: ...
\end{alltt}

\end{document}